\documentclass[pre,preprint,showpacs]{revtex4-1}
\usepackage[dvipdfmx]{graphicx}
\usepackage{amsmath,amssymb}
\usepackage{dcolumn}
\usepackage{bm}
\usepackage{hyperref}

\begin{document}
\title{Stationary state of a zero-range process corresponding to 
multifractal one-particle distribution}

\author{Hiroshi Miki}
\affiliation{Research Institute for Humanity and Nature,\\
457-4 Motoyama Kamigamo, Kita-ku, Kyoto, 603-8047, Japan}

\date{\today}

\begin{abstract}
We investigate a zero-range process where the underlying one-particle 
stationary distribution has multifractality. The multiparticle stationary 
probability measure can be written in a factorized form. If the number of 
the particles is sufficiently large, a great part of the particles condense 
at the site with the highest measure of the one-particle problem. 
The number of the particles out of the condensate scales algebraically 
with the system size and the exponent depends on the strength of the 
disorder. These results can be well reproduced by a branching process, 
with similar multifractal property.  
\end{abstract}

\maketitle

\section{Introduction}
It has been well known that disorder may change the behavior of the system, 
not only quantitatively but also qualitatively, in stochastic systems, 
as well as in classical or quantum spin systems and the quantum localization 
problem\cite{RBS,IM}. The effect of disorder is more remarkable when the 
spatial dimension of the system is lower. For example, at one-particle level, 
a localization where a particle stays at a certain specific site with much 
higher probability than at the other sites\cite{BG}, and a phase separation, 
where the regions with high particle density and low particle density are 
separated spontaneously\cite{Krug}, are well known phenomena. 

In many cases, the disorder which is considered is a random disorder: 
A disorder is represented as identically and independently distributed 
(i.i.d.) stochastic variables and thus, by definition, 
these variables are not spatially 
correlated each other. In a real situation, however, disorder may be 
spatially correlated and it is valuable to consider a model with spatially 
correlated disorder. 

Aperiodic disorder is a class of disorder with spatial correlation. 
An aperiodic disorder is generated deterministically  from a set of specific 
rules. That is the point which distinguishes an aperiodic disorder from 
random disorder and which enables us to study systems with aperiodic 
disorder systematically. Therefore study of systems with aperiodic disorder 
is a good first step toward understanding of more general system with 
spatially correlated disorder. Although some investigations have been carried 
out about the effect of aperiodic disorder in classical and quantum spin 
chains\cite{Luck,Hermisson,Vieira,Hida} so far, to our knowledge, 
just a few in stochastic 
systems. It has been known that even at one-particle level, an interesting 
behavior can be observed: An anomalous diffusion where the variance grows 
less slowly than linear growth in time \cite{ITR} and correspondingly the 
stationary probability distribution with multifractality\cite{Miki}. 
Note that systems with aperiodic disorder are not only of theoretical and 
mathematical interest, but also they have been artificially fabricated and 
investigated experimentally\cite{apexp}.

One of the next interesting problems is to make clear an interplay between 
quenched disorder with spatial correlation and interaction between particles. 
For this purpose we study a simple multiparticle stochastic process, 
called the zero-range process (ZRP)\cite{EH} with aperiodic disorder. 
In ZRP, it is remarkable that the exact stationary measure can be written 
in a factorized form, which is derived from the one-particle stationary 
probability distribution. Moreover, under a certain condition, condensation 
can be observed, where a great part of the particles condense at one 
specific site. The effect of random quenched disorder in ZRP has already 
been investigated\cite{JSI,GL}. In this paper, specifically, we consider 
ZRP with disorder constructed from an aperiodic sequence, 
called the paperfolding (PF) sequence\cite{Dekking}. 
This is one of the models where a multifractal stationary probability 
distribution can be observed.        

\section{A zero-range process with aperiodic disorder}
\begin{figure}
\begin{center}
\includegraphics[width=8cm]{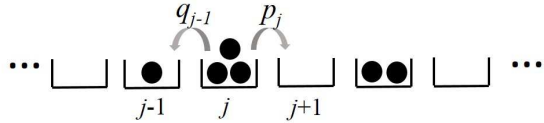}
\caption{ZRP on one-dimensional lattice with aperiodic disorder.
\label{zrp}}
\end{center}
\end{figure}

Let us consider a one-dimensional lattice with periodic boundary condition. 
Each site $j=1,2,\cdots, L$ can accommodate any number of particles. 
A particle hops from site to site with designated hop rates and only hopping 
to one of the nearest neighboring sites is allowed. The hop rates are quenched 
variables, where we denote the forward rate from site $j$ to $j+1$ by $p_j$ 
and the backward rate from $j$ to $j-1$ by $q_{j-1}$ (see Fig.\ref{zrp}).
Here we assume that the hop rates do not depend on the number of particles 
at the departure site (sometimes called the "pure chipping process"\cite{LMS}), 
although generally the hop rates in ZRP are given by a function of the number 
of particles at the departure site. For example, in a noninteracting particle 
system the hop rates are directly proportional to the number of particles 
at the departure site. 

Let us construct the aperiodic disorder according to the PF sequence. 
The PF sequence $S=AABAABBA\cdots$ is generated systematically 
by the initial sequence $S_1=AA$ and the substitution rules, 
$AA \rightarrow AABA$, $AB \rightarrow AABB$, $BA \rightarrow ABBA$, and 
$BB \rightarrow ABBB$. The length of the sequence of the $n$-th generation, 
$S_n$, is $2^n$ and in the $n \rightarrow \infty$ limit the ratio of the number 
of A's, $N_n(A)$, to that of B's, $N_n(B)$, converges to unity. 
Then the hop rates are assigned as
\begin{equation}
p_j=1,\quad \text{for all $j$},
\label{forwardrates}
\end{equation}   
and
\begin{equation}
q_j=
\begin{cases}
a, &\quad \text{the $j$-th symbol of $S$ is $A$},
\\
b, &\quad \text{the $j$-th symbol of $S$ is $B$}. 
\end{cases}
\label{backwardrates}
\end{equation} 

At one-particle level, the drift velocity $v_d$ is proportional to the 
difference between unity and the product of $q_j/p_j$ \cite{Derrida}
\begin{equation}
v_d \propto  1-\prod_{j=1}^L \frac{q_j}{p_j}.
\label{driftvel}
\end{equation}

The state of the multiparticle system is specified by the particle 
configuration 
$\{n_j\}=\{n_1,n_1,\cdots,n_L\}$, where $n_j$ denotes the number of particles 
at site $j$. The total number of particles, $N=\sum_{j=1}^Ln_j$ is conserved, 
since the periodic boundary condition is imposed. 

Let $P(\{n_j\};t)$ denote the probability that the configuration $\{n_j\}$ 
is observed at time $t$. The master equation is given as
\begin{eqnarray}
\frac{\partial}{\partial t} P(\{n_j\};t)
&=& \sum_j p_{j-1} P(\{\cdots, n_{j-1}+1,n_{j}-1,\cdots \};t)
\nonumber
\\
&&+\sum_j q_{j} P(\{\cdots, n_{j}-1,n_{j+1}+1,\cdots \};t)
\nonumber
\\
&&-\sum_j (p_{j}+q_{j-1})P(\{n_j\};t).
\label{mastereq}
\end{eqnarray}
One of the remarkable features of ZRP is that the exact multiparticle 
stationary measure  $P(\{n_j\})$ can be obtained, 
from the one-particle distribution $\{P_j\}$ 
(the stationary probability that the particle exist at the 
$j$-th site), in a factorized form:
\begin{equation}
P(\{n_j\})=\frac{1}{Z_{L,N}}\prod_{j=1}^L(P_j)^{n_j}\delta(\sum_{j=1}^Ln_j,N),
\label{prodmeasure}
\end{equation} 
where $Z_{L,N}$ denotes the partition function
\begin{equation}
Z_{L,N}=\sum_{\{n_j\}}\prod_{j=1}^L(P_j)^{n_j}\delta(\sum_{j=1}^Ln_j,N).
\label{partitionfunction}
\end{equation}
The symbol $\delta(\sum_{j=1}^Ln_j,N)$ indicates that the total number of 
particles is fixed at $N$. 

The one-particle distribution $\{P_j\}$ is given by the stationary solution 
of the master equation describing the random walk on the lattice:
\begin{equation}
\frac{\partial P_j(t)}{\partial t}
= p_{j-1}P_{j-1}(t) + q_{j}P_{j+1}(t)
-(p_{j}+q_{j-1})P_j(t).
\label{randomwalk}
\end{equation}
It is known that the stationary solution can be obtained 
analytically\cite{Derrida}:
\begin{equation}
P_j = \frac{r_j}{\sum_{k=1}^Lr_k},
\label{probdist}
\end{equation}
where  
\begin{equation}
r_j = \frac{1}{p_j}\left(1+
\sum_{i=1}^{L-1}\prod_{k=1}^i\frac{q_{j+k-1}}{p_{j+k}}\right).
\label{rj_aux}
\end{equation}

Under the condition for the drift velocity Eq.(\ref{driftvel}) to vanish
\begin{eqnarray}
b&=&a^{-\delta(n)},
\nonumber
\\
&&\delta(n)=N_n(A)/N_n(B),
\label{zerocurrent}
\end{eqnarray}
the stationary distribution $\{P_j\}$ shows multifractality and therefore 
the distribution is neither extended nor localized\cite{Miki}. 
Hereafter we restrict ourselves to this case. This model has only one 
free parameter $a$.

\section{Condensation and scaling of the number of particles out of condensate}
\begin{figure}
\begin{minipage}{0.8\hsize}
\begin{center}
\includegraphics[width=8cm]{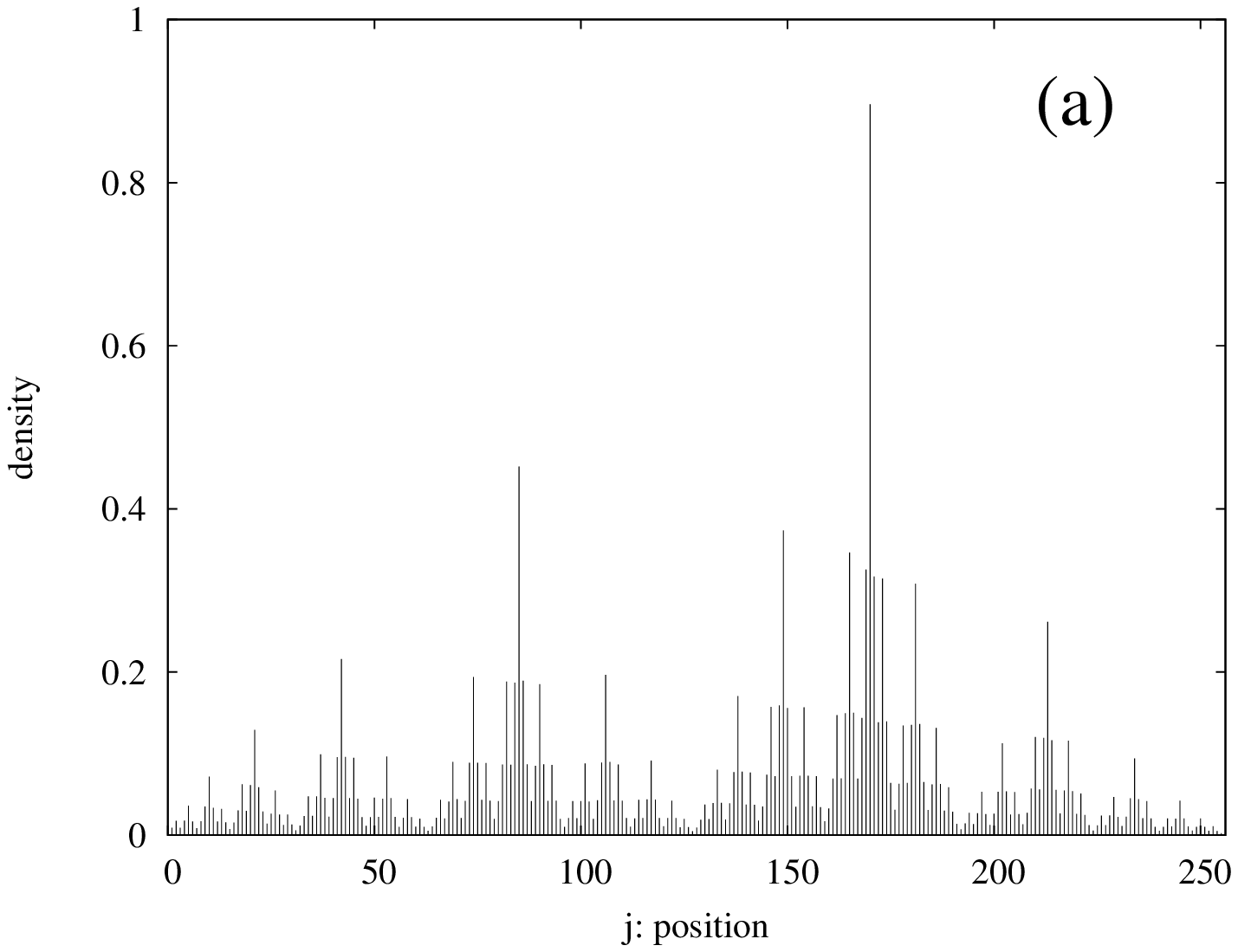}
\end{center}
\end{minipage}
\begin{minipage}{0.8\hsize}
\begin{center}
\includegraphics[width=8cm]{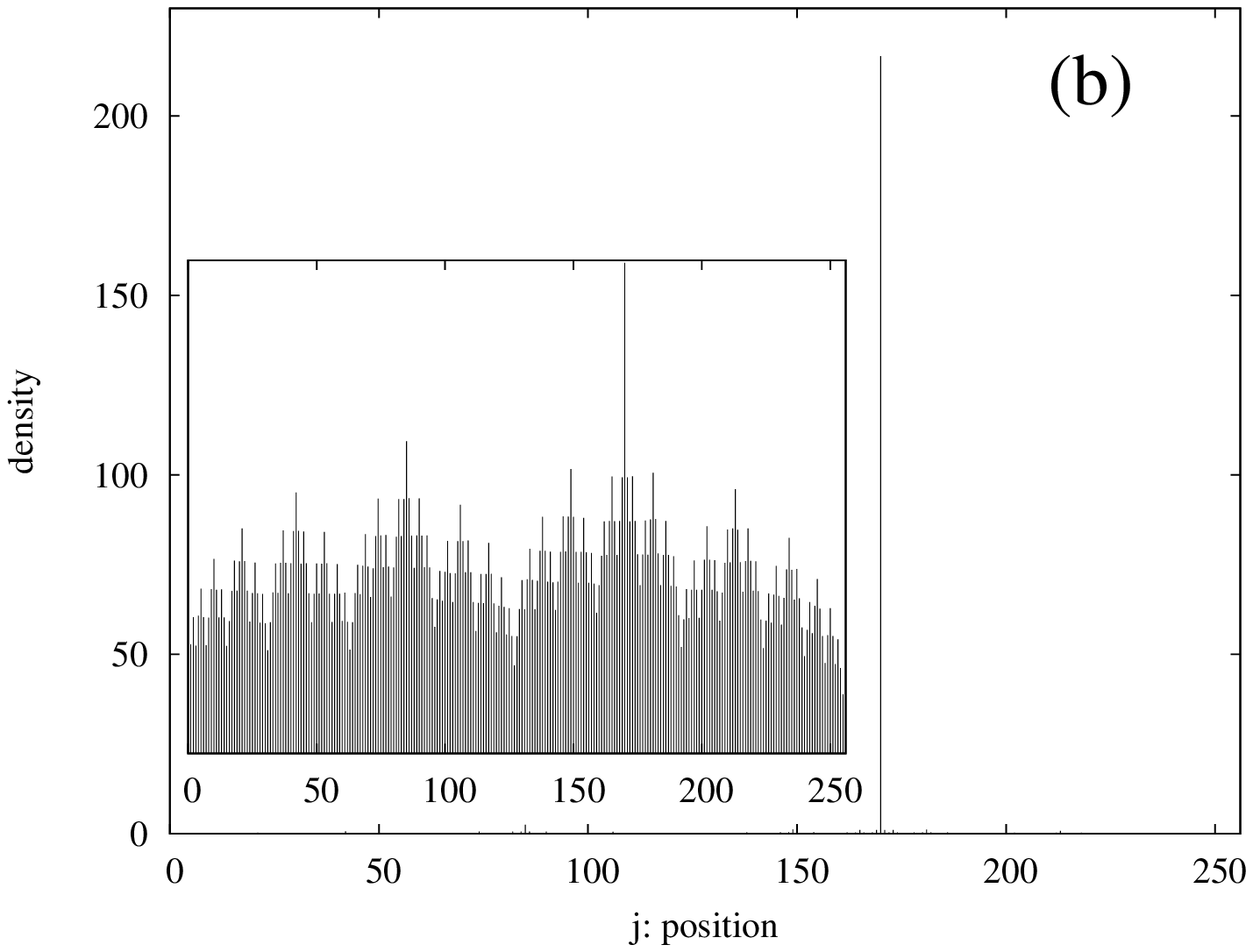}
\end{center}
\end{minipage}
\caption{Density profile obtained by simulations. (a) Low-density case, 
$L=256$ and $N=16$. (b) High-density case, $L=N=256$. The inset is the 
same plot with the ordinate in the log-scale. 
\label{profile}}
\end{figure}

Figure \ref{profile} shows the density profile obtained from Monte Carlo 
simulations.
When the average density is low, the interaction between particles is 
ineffective and the profile shows a hierarchical structure, as observed for
the one-particle distribution (see Fig.\ref{profile}(a)). 
On the other hand, when the average density increases, many particles 
condense at the site which gives the maximum of the one-particle 
distribution (see Fig.\ref{profile}(b)). 
This condensation is already known to occur for randomly disordered case. 
An analogy to the Bose-Einstein condensation is pointed out\cite{EH}.

In order to study the condensation in more detail, we introduce the 
grand-canonical(GC) formulation. In the GC formulation, the grand partition 
function ${\mathcal Z}_L(z)$ is given in a factorized form: 
\begin{eqnarray}
{\mathcal Z}_L(z) &=& \sum_{N \ge 0} z^N Z_{L,N}
\nonumber
\\
&=& \prod_{j=1}^L \frac{1}{1-zP_j},
\label{grandpartition}
\end{eqnarray}
where $z$ denotes the fugacity. Then the local density at site $j$ is 
obtained as 
\begin{equation}
\rho^{\rm GC}_j(z) = \frac{zP_j}{1-zP_j}.
\label{localdensity}
\end{equation} 
The average density is given as
\begin{equation}
\rho^{\rm GC}(z) = \frac{1}{L}\sum_{j=1}^L\frac{zP_j}{1-zP_j}.
\label{avdensity}
\end{equation}
From these expressions, it is immediately understood that the density of 
the site, at which the maximum of the one-particle distribution is given, 
rapidly increases, since at the site, as $z$ increases, the numerator of 
Eq.(\ref{localdensity}) increases most rapidly and the denominator 
decreases most rapidly. However, for a finite system, 
Eqs.(\ref{localdensity}) and (\ref{avdensity}) can take any finite value 
for $0 \le z <1/P_{\max}$. Therefore, the GC formulation cannot be broken 
as long as we consider a finite system. As is well known, condensation 
is characterized by the breaking of the GC formulation. Then it is necessary 
to consider the thermodynamic limit where $L,N \rightarrow \infty$ with 
$N/L$ fixed to be finite.  

In the thermodynamic limit, the expression of the total density 
Eq.(\ref{avdensity}) is rewritten as
\begin{eqnarray}
\rho^{\rm GC}(z) &=& \int_0^1dx\frac{zP(x)}{1-zP(x)},
\qquad(x:=j/L, L \rightarrow \infty) 
\nonumber
\\
&=& \int_0^{1/\max\{P_j\}}dpf(p)\frac{zp}{1-zp},
\label{TDdensity}
\end{eqnarray}
where the summation in the r.h.s. of Eq.(\ref{avdensity}) is replaced with 
the integral, and $f(p)$ denotes the distribution function of $p$. 
A condensation can occur if the integral converges in the limit 
$z \rightarrow 1/\max\{P_j\}$. For the integral to converge, the distribution 
function $f(p)$ must be an increasing function of $(p_{\max}-p)$ in the 
neighborhood of $p_{\max}=\max\{P_j\}$. In the case where the one-particle 
distribution $\{P_j\}$ has multifractality, this condition is considered 
to be satisfied. However, it is still quite 
difficult to calculate the integral and evaluate the critical density at 
which the condensation occurs. This consideration is similar to that 
for the density of states near the lowest energy level in Bose-Einstein 
condensation. Hereafter, we consider the case where the total density  
is large enough for the condensation to occur in the thermodynamic 
limit. 

The total number of particles out of the condensate, $N_{\rm out}$, is 
evaluated as 
\begin{eqnarray}
N_{\rm out} &=& \sum_{j \ne j_{\max}} \rho_j^{\rm GC}(z) 
\nonumber
\\
&=& \sum_{j \ne j_{\max}} \frac{zP_j}{1-zP_j}
\nonumber
\\
&\approx& \sum_{j \ne j_{\max}} \frac{P_j}{P_{\max}-P_j},
\label{nouteq}
\end{eqnarray}
where $j_{\max}$ is the condensation site and in the last line the fugacity 
$z$ is replaced with $1/P_{\max}$.

\begin{figure}
\begin{center}
\includegraphics[width=8cm]{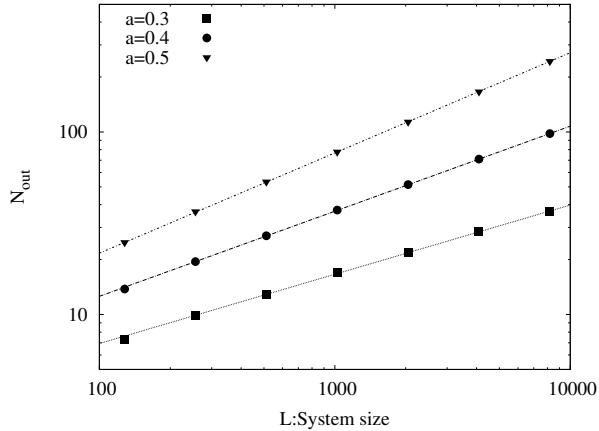}
\caption{System size dependence of the number of particles out of the 
condensate, $N_{\rm out}(L)$, for $a=0.3$, $0.4$, and $0.5$. The lines are 
guide for the eyes.
\label{nout}}
\end{center}
\end{figure}

Figure \ref{nout} shows the dependence of $N_{\rm out}$ 
on the system size $L$ for $a=0.3$, $0.4$, and $0.5$. It is found that 
$N_{\rm out}$ increases algebraically with respect to $L$:
\begin{equation}
N_{\rm out} \sim L^{\gamma(a)},
\label{noutpower}
\end{equation}
where the power-law exponent $\gamma$ depends on $a$. This result is 
qualitatively different from that in the case with random disorder, 
where $N_{\rm out}$ is kept ${\mathcal O}(1)$, independent of the 
system size\cite{JSI}. This can be considered as one of the 
characteristic results generated by the spatially 
correlated aperiodic disorder and interaction between particles. 

\begin{figure}
\begin{center}
\includegraphics[width=8cm]{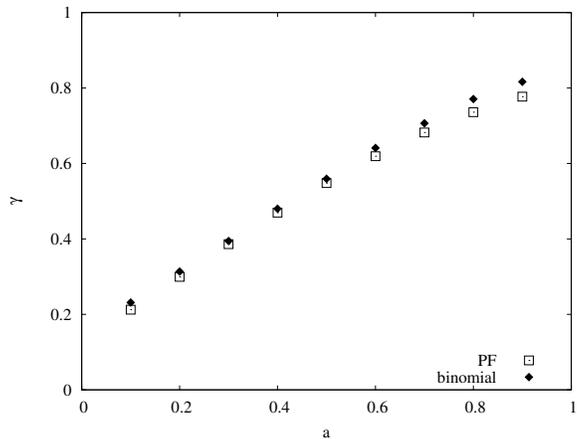}
\caption{Dependence of power-law exponent on disorder strength, 
$\gamma(a)$ in Eq.(\ref{noutpower}) for the PF model and the BBP.  
\label{gma}}
\end{center}
\end{figure}

Figure \ref{gma} shows the dependence of the exponent 
$\gamma$ on $a$. It is a monotonically increasing function: The stronger the 
disorder, the smaller the exponent. However, we are not sure about the values
to which the exponent converges in the $a \rightarrow 0$ (extremely strong 
disorder) and $a \rightarrow 1$ (extremely weak disorder) limits. It is quite 
difficult to obtain them numerically, since for $a \rightarrow 0$ some 
measures are quite small, and for $a \rightarrow 1$ the difference of the 
measures is quite small.  
 
In our previous study, we found that the multifractal structure of 
the one-particle distribution can be well reproduced by the binomial 
branching process (BBP)\cite{Miki}. This process is constructed by 
the iteration of dividing the segments into two halves and 
uneven partitioning of the measure by assiging one segment to $p$ and the 
other to $(1-p)$\cite{CMKRS}, where $p$, assumed to be $>1/2$, 
is the only free parameter of the process. 
The disorder strength of the PF model, $a$, and the parameter $p$ can be 
related through the multifractal singular exponent: 
\begin{equation}
\alpha_{\min}(a) = -\log_2p.
\label{pf_bbp}
\end{equation}
For the distribution generated by the BBP with $p$, which corresponds to 
the given disorder strength $a$, we evaluate $N_{\rm out}$ through 
Eq.(\ref{nouteq}). The evaluated $N_{\rm out}$ also shows a power-law 
dependence on the system size $L$, 
similar to that for the PF model. Moreover, the 
exponent $\gamma$ agrees very well, as shown in Fig.\ref{gma}. 
The one-particle distributions themselves for these two processes are 
not so similar each other, 
although their multifractal structures are similar. This suggests that 
the multifractal structure of the one-particle distributionis essential. 
It should be noted that for the PF model the structure is resulted from 
the aperiodic disorder.  
 
\section{Summary}
We investigated a phenomena emerged by both an aperiodic disorder and 
interaction between particles using the ZRP with disorder according to 
the PF sequence, for which the underlying one-particle stationary 
distribution is multifractal. In the stationary state, when the density 
is sufficiently high, a condensation occurs, where most particle 
condense at one specific site. For a finite size system, the number of 
particles out of the condensate increases with the system size 
algebraically, contrary to the randomly disordered case where it is 
suppressed to ${\mathcal O}(1)$. The distribution generated by the BBP 
has a similar multifractality and can reproduce the exponent well. 
This is a characteristic result by an interplay between the effects of 
spatially correlated disorder and that of the interaction between 
particles.

\begin{acknowledgments}
This research was supported by the initiative-based project E-05 
"Creation and Sustainable Governance of New Commons through Foundation of 
Integrated Local Environmental Knowledge (ILEK)", Research Institute for 
Humanity and Nature (RIHN).
\end{acknowledgments}

\end{document}